# Evaluation of microscale crystallinity modification induced by laser writing on Mn$_3$O$_4$ thin films

Camila Ianhez-Pereira[1], Akhil Kuriakose[2,3], Ariano De Giovanni Rodrigues[1], Ana Luiza Costa Silva[1], Ottavia Jedrkiewicz[2], Monica Bollani[2] and Marcio Peron Franco de Godoy[1]

[1]*Universidade Federal de São Carlos, Departamento de Física, Brazil*
[2] *Istituto di Fotonica e Nanotecnologie-Consiglio Nazionale delle Ricerche (IFN-CNR), Como, Italy*
[3]*Università dell'Insubria, Dipartimento di Scienza e Alta Tecnologia, Italy*

**Abstract**

Defining microstructures and managing local crystallinity allow the implementation of several functionalities in thin film technology. The use of ultrashort Bessel beams for bulk crystallinity modification has garnered considerable attention as a versatile technique for semiconductor materials, dielectrics, or metal oxide substrates. The aim of this work is the quantitative evaluation of the crystalline changes induced by ultrafast laser micromachining on manganese oxide thin films using micro-Raman spectroscopy. Pulsed Bessel beams featured by a 1 µm-sized central core are used to define structures with high spatial precision. The dispersion relation of Mn$_3$O$_4$ optical phonons is determined by considering the conjunction between X-ray diffraction characterization and the phonon localization model. The asymmetries in Raman spectra indicate phonon localization and enable a quantitative tool to determine the crystallite size at micrometer resolution. The results indicate that laser-writing is effective in modifying the low-crystallinity films locally, increasing crystallite sizes from ~8 nm up to 12 nm, and thus highlighting an interesting approach to evaluate laser-induced structural modifications on metal oxide thin films.

**Keywords**: oxides; laser writing, phonon localization, instrumentation.

## 1. INTRODUCTION

Manganese oxides are an intriguing class of materials tuned by the Mn element due to their many valence states [1]. Among many configurations, the coexistence of $Mn^{2+}$ and $Mn^{3+}$ stabilizes the oxide as $Mn_3O_4$. This compound presents the hausmannite phase, which is characterized by a tetragonal symmetry with lattice parameters a = 5.76 Å and c = 9.46 Å. In particular, $Mn_3O_4$ is featured by exciting properties related to its morphology, such as catalytic activity [1,2] and electrochemical properties, making it suitable for energy storage supercapacitor technology [3,4].

Additionally, current devices have harnessed the advantages of micro or even nanoscale features, as evidenced by the remarkable energy storage capabilities demonstrated by supercapacitors incorporating binary metal oxide nanostructures [5]. Notably, there are also flexible micro-supercapacitors based on manganese oxide that exhibit high performance [6]. Furthermore, the exceptional capabilities of lithium-ion batteries are prominently observed when they are constructed with manganese oxide nanocrystals as a key component [7].

Modern advanced micro/nano-fabrication technologies, ranging from electron beam lithography, advanced optical lithography, and 3D printings, have inspired the development of high-performance novel micro/nanophotonic devices with outstanding mechanical and optoelectronic properties for various sensing and detecting applications. For instance, microfabrication techniques effectively produce sophisticated etched structures such as waveguides, metasurfaces, and microdisks for optoelectronic devices [8-12].

Material modifications with a micrometer or nanometer resolution can be achieved by techniques such as ultrafast laser writing. Thanks to the nonlinear absorption process occurring when an intense laser pulse interacts with a transparent material, it is possible by using laser microfabrication, to induce local refractive index changes inside the bulk, or even cracks or voids [13,14]. For instance, femtosecond laser writing has emerged as a powerful technique, capable of writing light-guiding structures with 3D configurations as well as creating nitrogen-vacancy complexes in bulk materials such as glass or diamond, generating high contrast refractive index changes [15,16].

In general, ultrafast lasers with femtosecond or picosecond pulse durations offer the possibility of extremely high heating rates in the irradiated spots of the materials, without disturbing nearby areas. It is the rapid and localized energy deposition that enables precise laser-induced phase changes in the amorphous or crystalline structure of transparent materials [17].

The phase changes induced by the laser writing of different microstructures on the surface of a crystal (such as stripes, grids, and disk patterns) depend strongly on the initial sample crystallinity, which can be controlled during the growth processes. The most used technique to characterize crystallinity changes is X-ray diffraction (XRD) as it directly measures crystalline ordering and stress over films by evaluating crystallite size and shifts in Bragg diffraction angles [18,19]. The resolution typically available with this technique is in the order of millimeters. To get down to micrometric resolution, it is necessary to use XRD with energy beams produced by synchrotrons, allowing the spatial mapping of the lattice parameters parallel and perpendicular to the sample surface with a spatial resolution down to 100 nm [20]. In contrast, analyzing crystallinity properties in microstructures is not trivial and usually requires unfriendly and expensive techniques such as transmission electronic microscopy (TEM).

Through this work, we show that by using non-destructive μ-Raman (micro-Raman) spectroscopy, it is possible to evaluate local crystallinity changes in the microstructures fabricated by ultrafast laser writing in $Mn_3O_4$ thin films. Different manganese oxide ($Mn_3O_4$) films have been grown by spray pyrolysis with different molarities of the precursor solution. While the XRD technique is used to derive the crystallite dimensions of the films, a phonon localization model is applied by considering that the phonons are confined within the crystalline domains. From this, the dispersion relation of $Mn_3O_4$ optical phonons is estimated, which enables us to quantify structural differences induced by the interaction of the samples with the laser, by fitting the measured μ-Raman spectra with the theoretical model based on the $q \neq 0$ phonon contribution occurring in the inelastic scattering. Micro-Raman spectroscopy is therefore applied to the modified areas of the thin films and our results highlight the crystallinity modifications tuned by laser writing. Our data also indicates the range of laser writing parameters for which the micromachining has a significant role in the crystallinity changes of the $Mn_3O_4$ films, highlighting the absence of induced stress around the designed microstructures.

## 2. MATERIALS AND METHODS

### a. Sample Growth

Thin manganese oxide ($Mn_3O_4$) films with different molarities were grown by spray pyrolysis on soda-lime glass substrates. The technique involves atomizing an aqueous solution into microdroplets using dry compressed air as a carrier onto a heated substrate. Near the surface, the microdroplets undergo pyrolysis, resulting in solvent evaporation and the breakdown of solute molecules. The solute for $Mn_3O_4$ is the manganese acetate tetrahydrate precursor ($Mn(CH_3COO)_2 \cdot 4H_2O$). The pyrolysis process is similar to those employed for NiO [21] and ZnO [22] growth, in which the chemical reaction occurs in two steps. After the dehydration of manganese acetate,

$$Mn(CH_3COO)_2 \cdot 4H_2O \rightarrow Mn(CH_3COO)_2 + 4H_2O,$$

the breaking of organic molecules gives

$$6Mn(CH_3COO)_2 + 25O_2 \rightarrow 2Mn_3O_4 + 24CO_2 + 18H_2O,$$

leaving as clean products $CO_2$, $H_2O$ and the oxide film. The deposition is carried out by a cyclic process that starts at 300°C and is interrupted at 220°C due to the non-stability of the temperature, as described in [23]. The solution flow was set at 0.2 – 0.3 mL/min and the carrier gas at 0.1 MPa. The number of moles of solute per one liter of solution, also known as molarity (M=moles/L), plays a fundamental role in modulating the film crystallinity conditions [23,24]. In this work, we have obtained manganese oxide films with three different molarities, namely: $M_1 = 5.10^{-1}$ M, $M_2 = 5.10^{-2}$ M, and $M_3 = 5.10^{-3}$ M which provided a thickness of the films of around 6 µm, 1 µm and 0.7 µm, respectively.

### b. Laser micromachining

The laser microfabrication of the $Mn_3O_4$ thin films was performed employing a 20-Hz Ti:Sapphire amplified laser system (Amplitude) delivering 40-fs transform-limited pulses at 790 nm wavelength in the mJ pulse energy range. By detuning the laser compressor, the pulse duration can be adjusted, allowing us to work in both femtosecond and picosecond regimes. In this work, the peculiarity of the micromachining process was the use of a highly focused nondiffracting beam, notably the so-called Bessel beam (BB), having a transverse pattern characterized by an intense central core surrounded by weak rings (see inset of Fig.1), which act as the conical energy reservoir for the reconstruction of the central core during propagation [25]. The pulse energy used is thus distributed among the core and rings of the beam, but it should be noted that only the central peak of the BB interacts with the material through the nonlinear absorption process, guaranteeing the precision of the microfabrication thanks to the localized plasma generation. Here the Bessel beam used was obtained by reshaping the 5 mm diameter laser beam through a conical lens (178° apex angle). After a suitable demagnification utilizing a telescopic system consisting of a 300 mm focal length lens ($L_2$) and a 20X microscope objective (9 mm focal length), the Bessel beam was featured at the sample position by a 380 µm focal line and a narrow central spot of 1 µm size and was injected vertically (along z) to the deposited films (See Fig.1). Only the end portion of the focal line (much longer than the thickness of the manganese oxide films) was used to interact with the thin film, so to avoid damage of the internal glass substrate. A pulse duration of 1 ps was set to be able to increase the range of pulse energies leading to material modification without complete ablation.

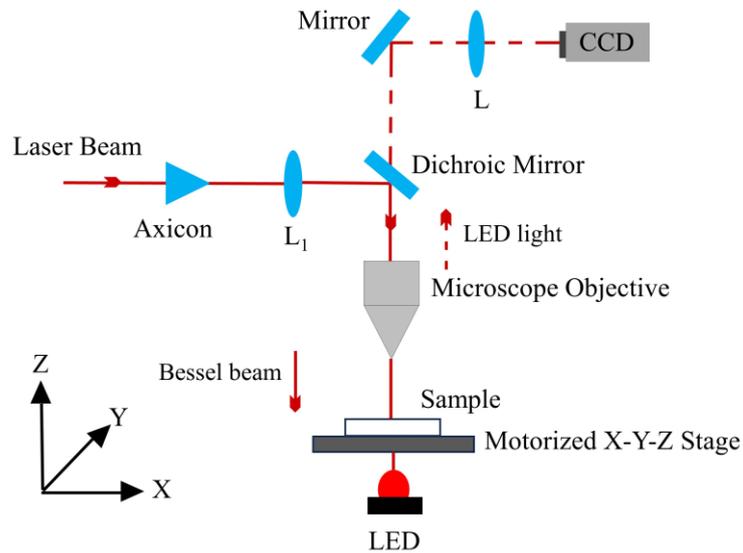

**Figure 1:** Scheme of the Bessel beam micromachining set-up used to laser write the manganese oxide films. In the inset, microscope image of the transverse pattern of the beam along the non-diffracting zone.

The ultra-high spatial localization of the Bessel beam core allowed us to precisely address the thin film crystalline modification. The pulse energy was set through a half-wave plate followed by a polarizer and the number of pulses delivered was controlled by using a mechanical shutter. The sample was positioned on a 3D motorized stage controlled via software, allowing to move the sample in three directions concerning the beam with a resolution of 100 nm. An imaging system (lens $L_1$ + objective) with LED illumination was set to image real-time the sample onto a CCD camera for Bessel beam machining, and in particular, for thin film microstructuring there was no need to move the sample along the beam (z) direction once the relative positioning between focal line and sample has been optimized. The surface micromachining following different types of design, such as thin lines (1 μm thick) disposed of in an array-type configuration, stripes (10 μm thick), or disks (20 or 40 μm diameter) was investigated by varying the pulse energy, and the writing speed. This has allowed us to observe in different regimes different types of material modifications. The microstructures obtained have then been analyzed by optical microscopy (OM) and scanning electron microscopy (SEM). Note that when writing lines, the Bessel beam machining was performed in one pass and the sample was shifted in one direction (x or y) in the horizontal plane, while for the stripes the writing lines were repeated (40 times) parallel to one another with a separation of 0.25 μm. For what concerns the disks, Bessel beam machining was performed along (20 or 40) circular trajectories with an increasing radius leading to 20 μm or 40 μm diameter respectively laser machined disk-type patterns on the sample surface.

### c. XRD and μ-Raman Characterization

X-ray diffraction measurements were performed with a Bruker X-ray diffractometer D8 Advance ECO configured in the Bragg-Brentano geometry over the scanning range 10°–70° and with the characteristic Cu-Kα source at 1.5406 Å. μ-Raman spectroscopy was realized employing a high-resolution spectrometer Horiba HR800 Evolution in combination with a continuous laser 633 nm as an excitation source. The μ-Raman spectra recorded with a 1.2 μm spatial resolution were featured by an uncertainty of about 0.7 cm$^{-1}$. All the measurements were taken with well-controlled laser power density to avoid heating effects.

## 3. RESULTS AND DISCUSSION

### 3a. X-Ray Diffraction (XRD)

Figure 2(a) shows the XRD patterns of the manganese oxide films $M_1 = 5 \times 10^{-1}$M, $M_2 = 5 \times 10^{-2}$M, and $M_3 = 5 \times 10^{-3}$M. According to the $Mn_3O_4$–reference from the International Crystal Structure Database (ICSD card 068174), the films are polycrystalline with the tetragonal hausmannite phase. The band centered at 25⁰ is due to the glass substrate which is enhanced for thinner films. After performing the background subtraction, a Gaussian fit of the peak corresponding to the (211) direction has been performed to evaluate its full width at half maximum (FWHM), as shown by the normalized curves of Fig. 2(b).

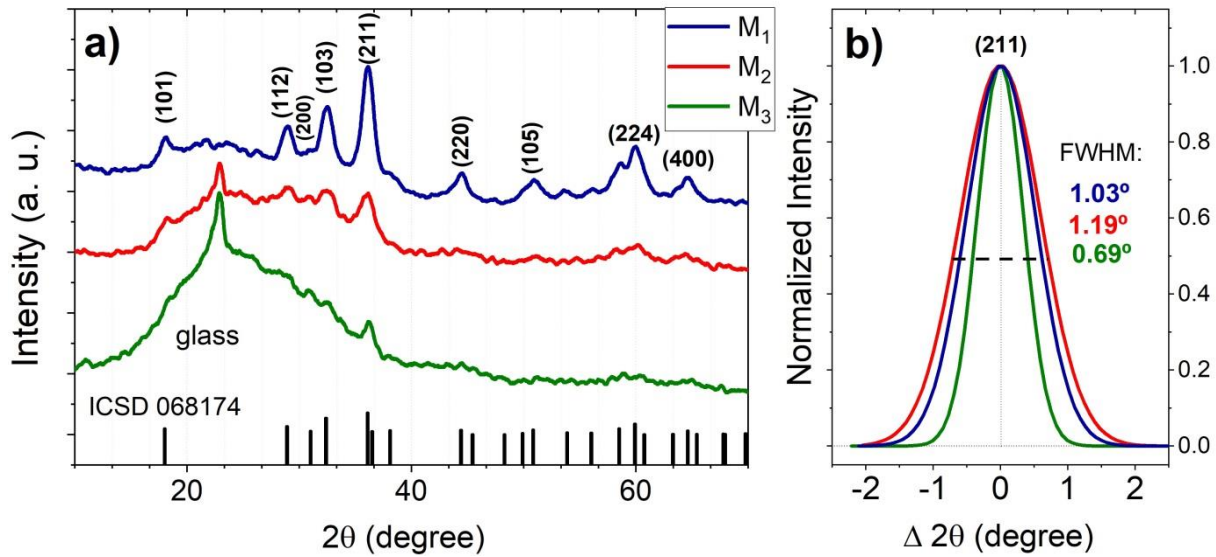

**Figure 2:** (a) XRD patterns of $Mn_3O_4$ thin films ($M_1$ = 5x10$^{-1}$ M, $M_2$ = 5x10$^{-2}$ M and $M_3$ = 5x10$^{-3}$ M) and the reference pattern (ICSD 068174) show a tetragonal hausmannite structure. (b) The normalized Gaussian fits for the (211) direction peak yielded FWHM values of 1.03⁰, 1.19⁰, and 0.69⁰ for films $M_1$, $M_2$, and $M_3$, respectively.

The crystalline ordering is quantified by the crystallite size *D* at one *(hkl)* direction making use of the Scherrer equation [26]:

$$D_{(hkl)} = \frac{0.9 \cdot \lambda}{FWHM_{hkl} \cdot \cos\theta_{hkl}} \quad (1)$$

where $\lambda$ is the X-ray wavelength, FWHM is the linewidth in radians, and $\theta_{hkl}$ is the peak position at the (*hkl*) direction.

Considering the peaks linewidth of the (211) direction, we estimate from Eq. (1) that the sample $M_1$ is featured by $D$ = (8.5 ± 1.0) nm, the sample $M_2$ by $D$ = (7.2 ± 1.0) nm and the sample $M_3$ by $D$ = (12.7 ± 0.8) nm as reported in the second column of Table 1. The difference in the crystallite size between the samples of higher molarities ($M_1$ and $M_2$) and that of lower molarity ($M_3$) is remarkable. The molarity is responsible for the concentration of metallic atoms that move on the neighborhood of the substrate surface until find a favorable condition to bind (or not) to an oxygen atom. Thus, Mn atoms have a significant mean free path to find the favorable energy to crystallize in films with lower molarities. However, the higher number of metallic atoms in the film $M_1$ = 5x10$^{-1}$M favors random crystallization like in a powder production process. This is also observed at a macroscopic scale: once synthesized, the film $M_3$ presents a compact film attached to the substrate, while the film $M_1$ is composed of tiny particles.

### 3b. Laser Writing and SEM morphological characterization

The microstructuring of the manganese oxide films, with patterns of different shapes such as thin lines, stripes, or disks, was realized by Bessel beam writing, varying different laser parameters such as pulse duration, pulse energy, and writing speed. Several tests were performed to find the most suitable writing parameters to obtain a localized structural modification without the complete ablation of the films, taking also into account the irregularity in the film thickness in some portions of the samples. In Figure 3, we present planar view SEM images depicting an array of micromachined lines in each of the $M_1$, $M_2$, and $M_3$ samples. These images are provided for illustrative and comparative purposes, aiming to examine the morphological differences resulting from the laser writing process. The notable surface modifications shown in the figure have been obtained in a single pass for a 1 ps pulse, and a Bessel pulse energy of (a) 3.4 µJ ($M_1$), (b) 1.1 µJ ($M_2$), and (c) 366 nJ ($M_3$). Notice that a decrease in pulse energy is needed not to ablate the film, as the molarity (and thus film thickness) decreases. In all cases, the writing speed was fixed at 0.005 mm/s, thus implying a partial superposition of 4 pulses per position.

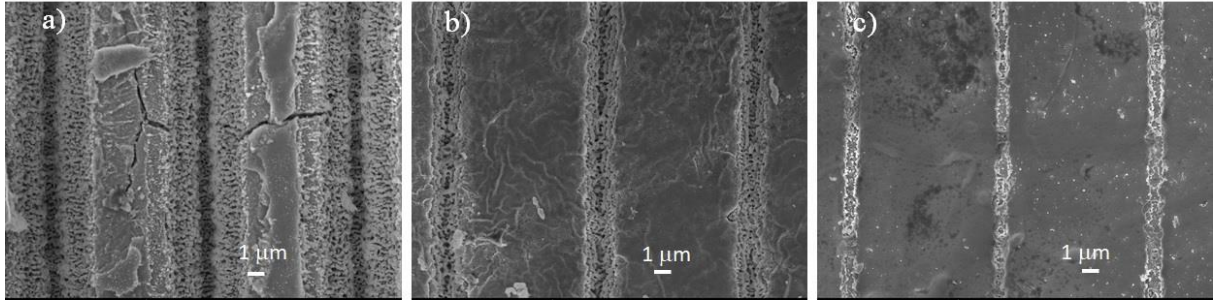

**Figure 3:** Planar view SEM images of aligned microstructures (array pattern) obtained by Bessel beam laser writing in the manganese oxide films (a) $M_1$, (b) $M_2$, and (c) $M_3$.

In Figure 4 we present planar view SEM images showing the stripes and disks-type patterns written by Bessel beam micromachining in the $Mn_3O_4$ thin films of different molarities. The pulse duration was 1 ps and the pulse energies used for these fabrications are respectively the same as those of Figure 3, respectively for the three types of films. The pulse duration was chosen to avoid a complete ablation of the film, but just above the threshold to observe under an optical microscope a material modification. Typically, as the molarity increased we noticed an increase in this energy threshold value. We also mention that the disks have been micromachined with a 0.01 mm/s writing speed instead of 0.005 mm/s as used for the lines and stripes, to reduce the number of superimposing pulses, since stronger ablation for disk patterns was observed, for the same energy or pulse durations. Also, note that the disks in Figs. 4(d) and 4(e) have been obtained with a smaller number of circular writing trajectories (20) than that shown in Fig. 4(f) (40), which explains the smaller sizes.

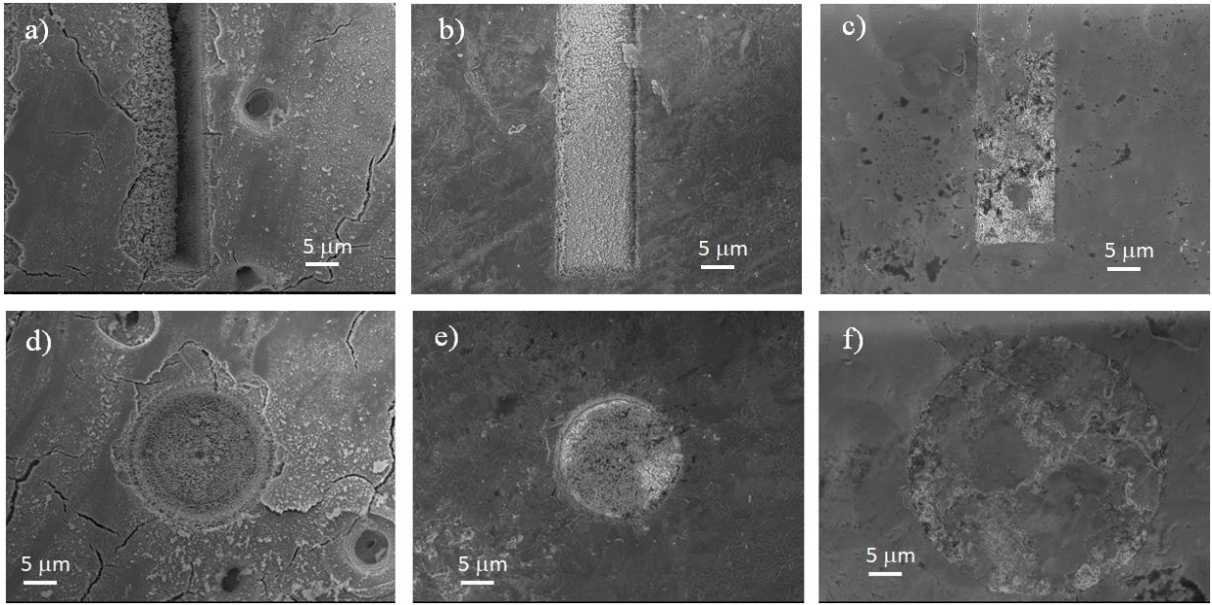

**Figure 4:** SEM images of stripes and disks-type patterns written by Bessel beam micromachining in the $Mn_3O_4$ thin films, (a) and (d) film $M_1$ (pulse energy 3.4 µJ); in (b) and (e) film $M_2$ (pulse energy 1.1 µJ); and in (c) and (f) $M_3$ (pulse energy 366 nJ). The writing speed was 0.005 mm/s for a), b), and c), and 0.01 mm/s for d), e), and f).

### 3c. Micro-Raman Spectroscopy and evaluation of the phonon localization length

The high sensibility of the µ-Raman line shape to the structural characteristics of solids indicates the possibility of using it for evaluating the microstructural variations. Figure 5 presents the typical µ-Raman spectra of the $Mn_3O_4$ films in the range of the optical phonons. The prominent peak is assigned to the scattering of optical phonons with $A_{1g}$ symmetry of the hausmannite structure [27,28]. In single crystals, it would be expected to find $A_{1g}$ peaks with a narrow and symmetric profile. However, when the crystalline domain decreases to nanometric dimensions, $A_{1g}$ assumes an asymmetric shape, with the FWHM increasing as the crystallite size decreases [29].

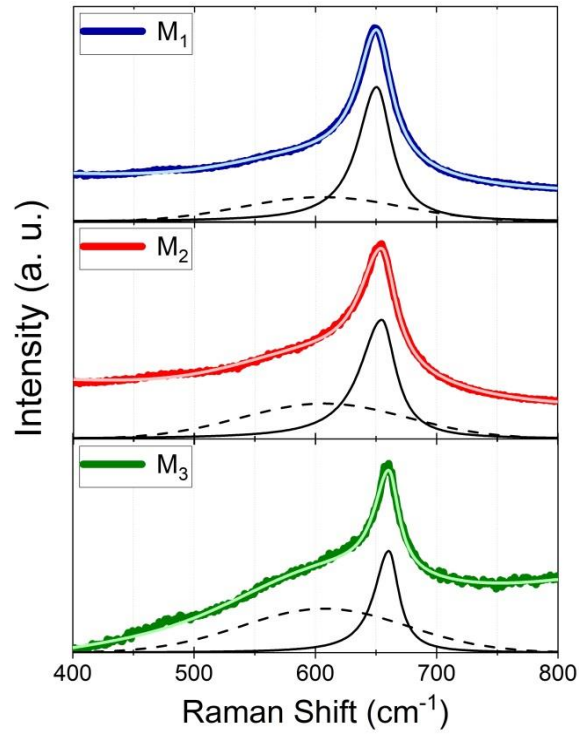

**Figure 5:** μ-Raman spectra of the $Mn_3O_4$ thin films ($M_1 = 5\times10^{-1}$M, $M_2 = 5\times10^{-2}$M and $M_3 = 5\times10^{-3}$M), and fits using Eq. (2) for the $A_{1g}$ peak (solid black line) and a Gaussian curve (dashed line) corresponding to a contribution activated by the disorder.

In fact, structural analyses of modification in the long-range ordering using the phonon localization model are well-established in the literature, especially for semiconductor materials [30-32]. Such a model accounts for the fact that, in the presence of structural defects or when dealing with polycrystalline matrices, the decrease in the crystalline coherence length causes the phonons to be localized within small portions restricted to the dimensions of the crystalline domains. Such localization of the phonons corresponds to the breaking of the μ-Raman selection rules for momentum $q = 0$, due to the increase in the momentum uncertainty. The contributions of phonons with $q \neq 0$ in the inelastic scattering of light result in a µ-Raman peak with a modified profile in comparison to that found in single crystals, described by the equation [33]:

$$I(\omega) \, \alpha \int \exp{(-q^2 L^2/4)} \, \frac{d^3 q}{[\omega - \omega(q)]^2 + (\Gamma/2)^2} \qquad (2)$$

where $I(\omega)$ is the spectral intensity as a function of frequency (or wavenumber), $\Gamma$ is the natural phonon damping, $\omega(q)$ is the phonon dispersion relation, and $L$ is phonon localization length (considering that the phonon is localized within a spherical region), which is determinative in the resulting peak shape. For instance, for a negative phonon dispersion relation, the activation of phonons with $q \neq 0$ having lower wavenumbers results in an asymmetrical broadening toward the left, which is larger for smaller values of $L$.

However, the calculation of the µ-Raman spectral intensity through Eq. (2) requires the dispersion relation $\omega(q)$ for $A_{1g}$ phonons that, to our knowledge, has not yet been reported in the literature. A reasonable way to empirically derive the $A_{1g}$ dispersion relation is by considering that the phonons must be limited within spatial regions defined by crystalline domains and, consequently, the phonon localization length must be similar to the crystallite size $D_{(211)}$ in the polycrystalline films, which can be calculated through eq. (1). With a fixed $L$ value, $\omega(q)$ is the only adjustable parameter when using the theoretical Eq. (2) to fit the experimental µ-Raman spectra of the $Mn_3O_4$ films shown in Fig.5. From the fit of the three spectra, each one with its specific $L$ taken to be equal to the crystallite size of the film, we have determined the common dispersion relation for $A_{1g}$ as $\omega(q) = \omega_0 - 200q^2$, where $\omega_0$ is the frequency of the phonon at the Brillouin zone center $(q = 0)$. Note that in such calculations, we have used for the optical phonon the isotropic form approximation.

For all the data shown in Fig. 5, a Gaussian spectral component is fitting the left side contribution of the asymmetrical peak calculated with eq. 2. This feature corresponds to a disorder-activated spectral contribution, which is more intense when the crystallinity is lower. Such a disorder band has already been reported in the literature for the same type of system presenting tiny crystallite sizes [27,32] and is commonly found in materials like silicon presenting some degree of the disorder [35,36].

Once the dispersion relation for $A_{1g}$ has been determined by using eq. 2, from the fit of the Raman spectra measured in the laser-written regions of the films, we could derive the modified phonon coherence lengths, which is a measure of the structural modifications induced by the laser-matter interaction. Figure 6 shows the results of the µ-Raman spectral measurements recorded from the three different laser-written patterns fabricated on the three films with pulse energy of 3.4µJ ($M_1$), 1µJ ($M_2$), and 366nM ($M_3$) respectively.

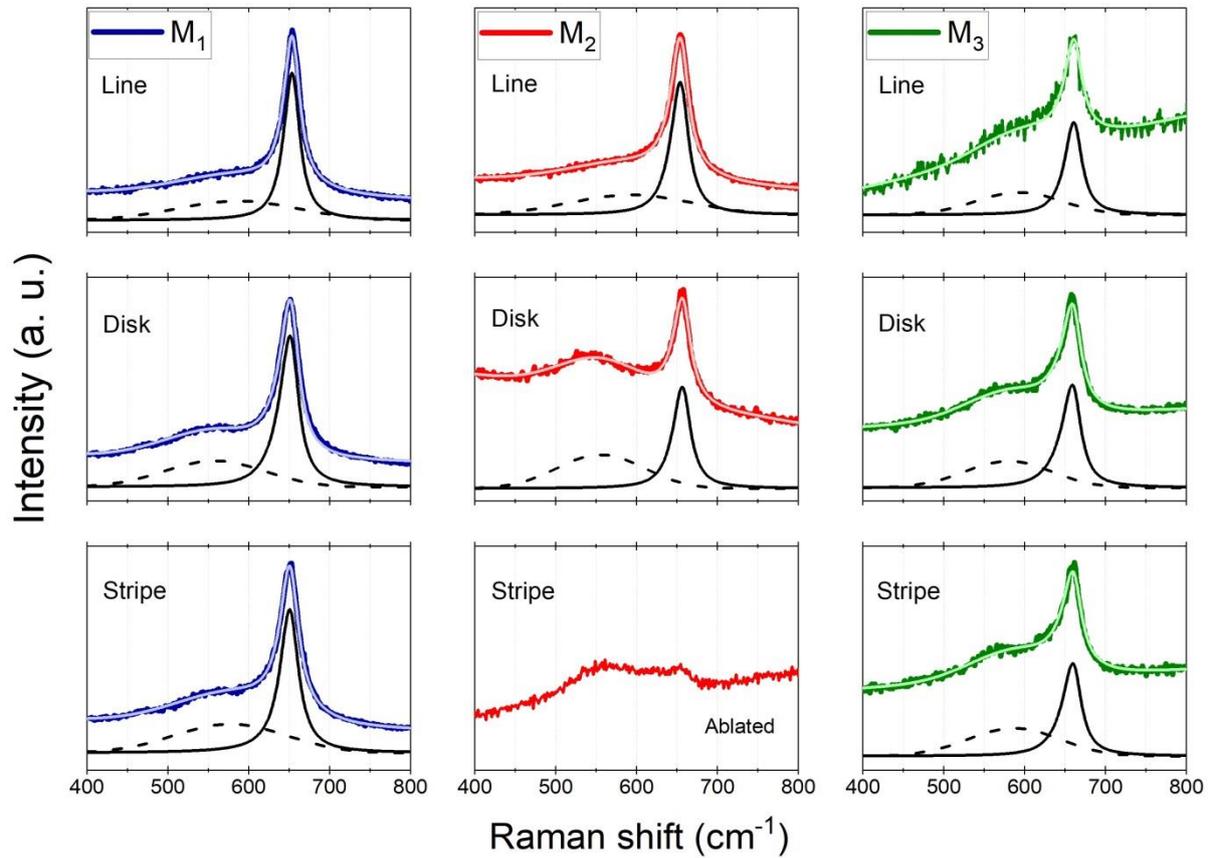

**Figure 6:** Spectra obtained from the μ-Raman measurements on $Mn_3O_4$ films ($M_1$, $M_2$, and $M_3$) performed on the line, disk, and stripe patterns. The fits of the different contributions are shown in a solid black line ($A_{1g}$ peak) and a dashed black line (disorder contribution), respectively.

The results obtained for the derivation of the phonon localization length $L$ in the three films analyzed have been reported in Table 1, together with the crystallite size calculated through eq.(1) retrieved from the XRD spectra. We notice how for the two samples grown in conditions of higher molarity, there is an increase in the crystallinity of the material when the structure undergoes modification due to the Laser-thin film interaction.

|  |  | Phonon location length (L) | | | |
| --- | --- | --- | --- | --- | --- |
| Molarity | $D_{211}$ (nm) | Film (nm) | Line (nm) | Disk (nm) | Stripe (nm) |
| $M_1 = 5 \times 10^{-1}$ M | 8.5 | 7.64 | 12.21 | 8.51 | 9.15 |
| $M_2 = 5 \times 10^{-2}$ M | 7.2 | 6.45 | 11.18 | 11.82 | ablated |
| $M_3 = 5 \times 10^{-3}$ M | 12.7 | 12.25 | 13.97 | 8.84 | 10.06 |

**Table 1:** Films crystallite sizes $(D_{211})$ derived by XRD and by using Eq. (1) and Phonon localization length $(L)$ extracted from Raman measurements performed on different laser written patterns and from Eq. (2), for different molarity films ($M_1 = 5 \times 10^{-1}$M, $M_2 = 5 \times 10^{-2}$M, $M_3 = 5 \times 10^{-3}$M).

The difference in the values of the phonon localization length before and after laser writing is found to be greater in the lines of the laser-written array pattern, where $\Delta L = 4.57\ nm$ for the film $M_3$ and $\Delta L = 4.73\ nm$ for the film $M_2$. In the case of the low molarity film $M_3$, the laser writing process did not generally lead to an increase in the crystalline correlation, contrary to what was observed for the films prepared with higher molarities. This indicates a limit to the improvement of $Mn_3O_4$ crystallinity by laser writing, which may depend on the initial structural characteristics of the film. Interestingly, the phonon localization length values evaluated by fitting the phonon contribution from the inelastic scattering in the Raman spectra together with the use of Eq.(2) are always slightly below the calculated crystallite size for the same sample. This result is expected given the fact that the phonons are confined within the perfect crystalline portions, so their movement is limited by the edges of the crystal which generally contain defects and dislocations.

Since Raman spectra peak positions are susceptible to stress/strain variation, comparative μ-Raman measurements were also performed at the edges of the laser-written patterns and on the original film (away from the edges) in all the micromachined configurations. Figure 7 shows an illustrative µ-Raman spectrum for the film $M_3$ at the edge of a micromachined disk pattern and on the film, with the respective optical microscope images to indicate the position of the laser beam used in the measurements in the two different cases.

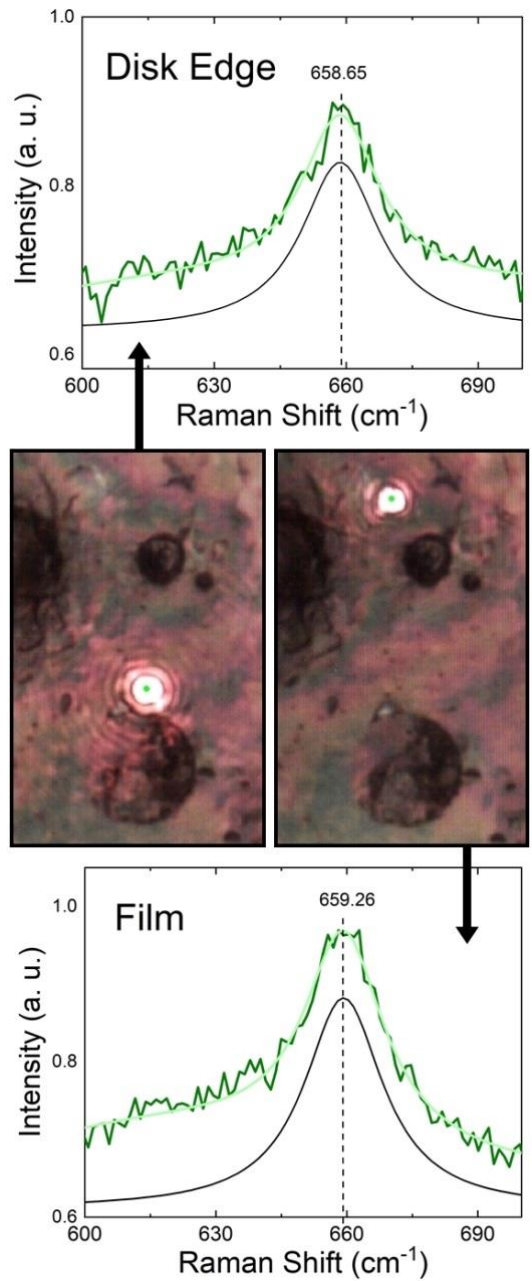

**Figure 7:** μ-Raman spectra relative to M$_3$ performed on the laser-written disk edge and on the film, in order to identify possible strains in the material due to laser writing. The optical images show the positioning of the bright laser spot used for the measurements.

Table 2 summarizes all the frequency shifts (peak position values of the spectra) obtained by the Lorentzian fit of the peak $A_{1g}$ at the edge of the laser writings and on the film.

| Molarity | Pattern | Edge (cm$^{-1}$) | Film (cm$^{-1}$) | Difference (cm$^{-1}$) |
|---|---|---|---|---|
| $M_1 = 5\times10^{-1}$ M | Line | 655.23 | 655.39 | 0.16 |
| | Disk | 656.73 | 656.25 | 0.48 |
| | Stripe | 654.29 | 654.77 | 0.48 |
| $M_2 = 5\times10^{-2}$ M | Line | 656.91 | 657.44 | 0.53 |
| | Disk | 657.14 | 656.98 | 0.16 |
| $M_3 = 5\times10^{-3}$ M | Line | 658.81 | 659.44 | 0.63 |
| | Disk | 658.65 | 659.26 | 0.61 |
| | Stripe | 658.78 | 659.16 | 0.38 |

**Table 2:** Peak position values (cm$^{-1}$) of the µ-Raman spectra recorded on all the films/ patterns selected and the difference between the peak positions measured on the edges of the laser-written patterns and the film.

The peak position variations shown in Table 2 are smaller than the experimental uncertainty of the measurement which is 0.7 cm$^{-1}$, confirming that no measurable strain difference between the film and the edges of the laser-written patterns can be detected. This indicates that the heating effect generated by the laser (picosecond Besel beam) is highly localized in the writing regions, and is not generating structural changes in the surroundings.

## 4. CONCLUSION

To summarize, we have shown through this work that the crystallinity of the $Mn_3O_4$ films can by modified through laser micromachining. The laser microfabrication was performed using ultrafast Bessel beams in the picosecond regime, featuring a non-diffracting core of 1 μm in size.

Our study on the crystallinity modification of the fabricated samples shows a dependence of the same on two factors, namely the molarity used in the solution and the type of laser written patterns analyzed, the latter mainly differing (given fixed laser parameters) by the type of writing trajectory and also by the number of superposing pulses at one precise location.

In our case, the results obtained on thin micrometer-sized lines, disks, or stripe patterns fabricated by the laser were slightly different. A notable change was found in the higher molarity films ($M_1$ and $M_2$) when comparing the phonon localization length before and after laser micromachining ($\Delta L \approx 4.6$ nm) for the line-type writing. This improvement is much smaller and not conclusive in the film $M_3$ ($\Delta L \approx 1.7$ nm). In this case, we observed a deterioration of crystallinity in disk and stripe-type laser written patterns, probably due to uncontrolled ablation effects resulting from the extremely low thickness of the film. In addition, the XRD analysis revealed that this film exhibited a higher crystallite size, indicating a potential limitation in enhancing the crystallinity through laser writing.

The method used here for calculating the phonon localization length proved to be very sensitive for the characterization of the crystalline properties of the $Mn_3O_4$ films. This sensitivity is evident from the excellent agreement observed between the calculated values of the localization length for the original films and the crystallite size determined from conventional XRD analyses. The use of this method enabled the quantitative evaluation of the changes in the crystallinity induced by ultrafast laser writing with a spatial resolution of a few micrometers. Such analysis would be unfeasible using conventional diffractometers based on X-ray techniques.

Finally, none of the films showed any trace of strain or stress at the edges of the microstructures fabricated using the laser. As changes in crystallinity have a consequent influence on the refractive index, we believe that laser microstructured $Mn_3O_4$ films have potential applications for waveguides beyond other optoelectronic devices with local photocatalytic activity and photoactivated sensing properties, and charge storage devices as supercapacitors.

## Acknowledgements

This research has received funding from European Union's H2020 Marie Curie ITN project LasIonDef (GA n.956387) and National Council for Scientific and Technological Development – CNPq - Brazil (311507/2020-4 and 309230/2020-9 ). ADR is grateful to São Paulo Research Foundation (FAPESP 2021/13974-0), and CIP and ALCS acknowledge their doctorate fellowships from CNPq and CAPES, respectively.

## REFERENCES

[1]   Y.F. Han, F. Chen, Z. Zhong, K. Ramesh, L. Chen, E. Widjaja, Controlled synthesis, characterization, and catalytic properties of $Mn_2O_3$ and $Mn_3O_4$ nanoparticles supported on mesoporous silica SBA-15, Journal of Physical Chemistry B. 110 (2006) 24450–24456. https://doi.org/10.1021/JP064941V

[2]   J.K. Pulleri, S.K. Singh, D. Yearwar, G. Saravanan, A.S. Al-Fatesh, N.K. Labhasetwar, Morphology Dependent Catalytic Activity of $Mn_3O_4$ for Complete Oxidation of Toluene and Carbon Monoxide, Catal Letters. 151 (2021) 172–183. https://doi.org/10.1007/S10562-020-03278-W.

[3]   D.P. Dubal, D.S. Dhawale, R.R. Salunkhe, S.M. Pawar, C.D. Lokhande, A novel chemical synthesis and characterization of $Mn_3O_4$ thin films for supercapacitor application, Appl Surf Sci. 256 (2010) 4411–4416. https://doi.org/10.1016/J.APSUSC.2009.12.057.

[4]   S.A. Beknalkar, A.M. Teli, T.S. Bhat, K.K. Pawar, S.S. Patil, N.S. Harale, J.C. Shin, P.S. Patil, $Mn_3O_4$ based materials for electrochemical supercapacitors: Basic principles, charge storage mechanism, progress, and perspectives, J Mater Sci Technol. 130 (2022) 227–248. https://doi.org/10.1016/J.JMST.2022.03.036.

[5]   A. M. Teli, T. S. Bhat, S. A. Beknalkar, S. M. Mane, L. S. Chaudhary, D. S. Patil, S. A. Pawar, H. Efstathiadis, J. C. Shin, Bismuth manganese oxide based electrodes for asymmetric coin cell supercapacitor, Chemical Engineering Journal, 430 (2022) 133138. https://doi.org/10.1016/j.cej.2021.133138.

[6]   X. Shi, Z. Zeng, C. Liao, S. Tao, E. Guo, X. Long, X. Wang, D. Deng, Y. Dai, Flexible, planar integratable and all-solid-state micro-supercapacitors based on nanoporous gold/ manganese oxide hybrid electrodes via template plasma etching method, J. Alloys Compd., 739 (2018) 979-986. https://doi.org/10.1016/j.jallcom.2017.12.292.

[7]   S. Chen, Y. Shi, Y. Wang, Y. Shang, W. Xiae and H. Y. Yang, An all manganese-based oxide nanocrystal cathode and anode for high performance lithium-ion full cells, Nanoscale Advances, 1 (2019) 1714-172. https://doi.org/10.1039/C9NA00003H.

[8]   G. Zhu, J. Li, N. Zhang, X. Li, J. Dai, Q. Cui, Q. Song, C. Xu, Y. Wang, Whispering-Gallery Mode Lasing in a Floating GaN Microdisk with a Vertical Slit, Sci Rep 10, 253 (2020) 1–7. https://doi.org/10.1038/s41598-019-57118-y.

[9]   H. Merbouche, B. Divinskiy, K. O. Nikolaev, C. Kaspar, W. H. P. Pernice, D. Gouéré, R. Lebrun, V. Cros, J. Ben Youssef, P. Bortolotti, A. Anane, S. O. Demokritov, V. E. Demidov, Giant nonlinear self-phase modulation of large-amplitude spin waves in microscopic YIG waveguides, Sci Rep., 12 (2022) 7246. https://doi.org/10.1038/s41598-022-10822.

[10] O. Jamadi, F. Reveret, P. Disseix, F. Medard, J. Leymarie, A. Moreau, D. Solnyshkov, C. Deparis, M. Leroux, E. Cambril, S. Bouchoule, J. Zuniga-Perez & G. Malpuech, Edge-emitting polariton laser and amplifier based on a ZnO waveguide, Light Sci Appl., 7 (2018) 82. https://doi.org/10.1038/s41377-018-0084-z

[11] N. Zhang, Y. Wang, W. Sun, S. Liu, C. Huang, X. Jiang, M. Xiao, S. Xiao and Q. Song, High-Q and highly reproducible microdisks and microlasers, Nanoscale, 10 (2018) 2045-2051. https://doi.org/10.1039/C7NR08600H.

[12] L. Fagiani, M. Gandolfi, L. Carletti, C. de Angelis, J. Osmond and M. Bollani, Modelling and nanofabrication of chiral dielectric metasurfaces, Micro and Nano Engineering, 19 (2023), 100187. https://doi.org/10.1016/j.mne.2023.100187

[13] R.R. Gattass, E. Mazur, Femtosecond laser micromachining in transparent materials, Nature Photonics, 2 (2008) 219–225. https://doi.org/10.1038/nphoton.2008.47

[14] O. Jedrkiewicz, S. Kumar, B. Sotillo, M. Bollani, A. Chiappini, M. Ferrari, R. Ramponi, P. D. Trapani, S. M. Eaton, Pulsed Bessel beam-induced microchannels on a diamond surface for versatile microfluidic and sensing applications, Optical Materials Express 7, 6 (2017) 1962-1970. https://doi.org/10.1364/OME.7.001962

[15] A. N. Giakoumaki, G. Coccia, V. Bharadwaj, J. P. Hadden, A. J. Bennett, B. Sotillo, R. Yoshizaki, P. Olivero, O. Jedrkiewicz, R. Ramponi, S. M. Pietralunga, M. Bollani, A. Bifone, P. E. Barclay, A. Kubanek, S. M. Eaton, Quantum technologies in diamond enabled by laser processing, Appl. Phys. Lett., 120, (2022) 020502. https://doi.org/10.1063/5.0080348

[16] S.M. Eaton, M.L. Ng, R. Osellame, P.R. Herman, High refractive index contrast in fused silica waveguides by tightly focused, high-repetition rate femtosecond laser, J Non Cryst Solids, 357 (2011) 2387–2391. https://doi.org/10.1016/J.JNONCRYSOL.2010.11.082.

[17] VV Belloni, M Bollani, SM Eaton, P Di Trapani, O Jedrkiewicz, Micro-hole generation by high-energy pulsed bessel beams in different transparent materials, Micromachines 12 (4), 455 https://doi.org/10.3390/mi12040455

[18]   V. Mondiali, M. Bollani, S. Cecchi, M. Richard, T. Schülli, G. Chahine, D. Chrastin, Dislocation engineering in SiGe on periodic and aperiodic Si(001) templates studied by fast scanning X-ray nanodiffraction, Applied Physics Letters, 104 (2014) 021918. https://doi.org/10.1063/1.4862688

[19]   G. Balakrishnana, R. Velavana, K. Mujasam Batoob, E. H. Raslanc, Microstructure, optical and photocatalytic properties of MgO nanoparticles, Results in Physics, 16 (2020) 103013 https://doi.org/10.1016/j.rinp.2020.103013.

[20]   D. Chrastina, G. M. Vanacore, M. Bollani, P. Boye, S. Schöder, M. Burghammer, R. Sordan, G. Isella, M. Zani, and A. Tagliaferri, Patterning-Induced Strain Relief in Single Lithographic SiGe Nanostructures Studied by Nanobeam X-ray Diffraction, Nanotechnology 23 (15) 155702(2012) https://doi.org/10.1088/0957-4484/23/15/155702

[21]   T. Fukui, S. Ohara, M. Naito, K. Nogi, Synthesis of NiO–YSZ composite particles for an electrode of solid oxide fuel cells by spray pyrolysis, Powder Technol. 132 (2003) 52–56. https://doi.org/10.1016/S0032-5910(03)00044-5.

[22]   Y. J. Onofre, S. de Castro, A. D. Rodrigues, M. P. F. de Godoy, Influence of Co-doping on optical properties and traps localization of ZnO films obtained by spray pyrolysis, J Anal Appl Pyrolysis. 128 (2017) 131–135. https://doi.org/10.1016/J.JAAP.2017.10.017.

[23]   C. Ianhez-Pereira, A. D. Rodrigues, M. P. F. de Godoy, Tailoring Stark effect in the1.54 μm emission of Er-doped ZnO thin films, Scr Mater. 192 (2021) 102–105. https://doi.org/10.1016/J.SCRIPTAMAT.2020.10.013.

[24]   M. P. F. de Godoy, L. K. S. de Herval, A. A. C. Cotta, Y. J. Onofre, W. A. A. Macedo, ZnO thin films design: the role of precursor molarity in the spray pyrolysis process, Journal of Materials Science: Materials in Electronics. 31 (2020) 17269–17280. https://doi.org/10.1007/S10854-020-04281-Y/ .

[25]   J. Durnin, J. J. Miceli, Jr., J. H. Eberly, Diffraction-free beams, Phys. Rev. Lett., 58 (1987) 1499. https://doi.org/10.1103/PhysRevLett.58.1499

[26] B. D. Cullity, J.W. Weymouth, Elements of X-Ray Diffraction, Am J Phys., 25 (2005) 394. https://doi.org/10.1119/1.1934486.

[27] H. D. Lutz, B. Müller, H. J. Steiner, Lattice vibration spectra. LIX. Single crystal infrared and Raman studies of spinel type oxides, J Solid State Chem. 90 (1991) 54–60. https://doi.org/10.1016/0022-4596(91)90171-D.

[28] S. Bernardini, F. Bellatreccia, G. D. Ventura, P. Ballirano and A. Sodo, Raman spectroscopy and laser-induced degradation of groutellite and ramsdellite, two cathode materials of technological interest, RSC Adv., 10 (2020) 923. https://doi.org/10.1039/C9RA08662E

[29] J. Zuo, C. Xu, Y. Liu, Y. Qian, Crystallite size effects on the Raman spectra of $Mn_3O_4$, Nanostructured Materials, 10 (1998) 1331–1335. https://doi.org/10.1016/S0965-9773(99)00002-1.

[30] A. D. Rodrigues, M. P. F. de Godoy, C. Mietze, D. J. As, Phonon localization in cubic GaN/AlN superlattices, Solid State Commun, 186 (2014) 18–22. https://doi.org/10.1016/J.SSC.2014.01.012.

[31] K. W. Adu, H. R. Gutiérrez, U. J. Kim, G. U. Sumanasekera, P. C. Eklund, Confined phonons in Si nanowires, Nano Lett. 5 (2005) 409–414. https://doi.org/10.1021/NL0486259/.

[32] K. K. Tiong, P. M. Amirtharaj, F. H. Pollak, D. E. Aspnes, Effects of As+ ion implantation on the Raman spectra of GaAs: '"Spatial correlation"' interpretation, Appl Phys Lett. 44 (1998) 122. https://doi.org/10.1063/1.94541.

[33] I. H. Campbell, P. M. Fauchet, The effects of microcrystal size and shape on the one phonon Raman spectra of crystalline semiconductors, Solid State Commun. 58 (1986) 739–741. https://doi.org/10.1016/0038-1098(86)90513-2.

[34] C. M. Julien, M. Massot, Raman spectroscopic studies of lithium manganates with spinel structure, Journal of Physics: Condensed Matter. 15 (2003) 3151. https://doi.org/10.1088/0953-8984/15/19/315.

[35]  B. Li, D. Yu, S. Zhang, Raman spectral study of silicon nanowires, Phys Rev B. 59 (1999) 1645. https://doi.org/10.1103/PhysRevB.59.1645.

[36]  Z. Iqbal, S. Veprek, Raman scattering from hydrogenated microcrystalline and amorphous silicon, Journal of Physics C: Solid State Physics. 15 (1982) 377–392. https://doi.org/10.1088/0022-3719/15/2/019.